\def\al26{\mbox{$^{26}$\hspace{-0.2em}Al}}          
\def\MeV{\mbox{Me\hspace{-0.1em}V}}                 
\def\keV{\mbox{ke\hspace{-0.1em}V}}                 
\def\deg{\mbox{$^\circ$}}                           
\def\dg{^\circ}                                     
\def\lb{{($l$, $b$)}}                               
\def\phibar{$\bar{\varphi}$}                        
\def\Msol{\mbox{M$_{\odot}$}}                       
\def\HI{\mbox{H\hspace{0.2em}{\scriptsize I}}}      
\def\HII{\mbox{H\hspace{0.2em}{\scriptsize II}}}    
\def\etal{{et al.}}                                 
\def\gray{\mbox{$\gamma$-ray}}                      
\def\reference{}
\def\AA{A\&A}                                       
\def\ApJ{ApJ}                                       
\def\ApJS{ApJS}                                     

\documentstyle[psfig]{l-aa}

\begin{document}

\thesaurus{04(13.07.2; 02.14,1; 10.19.3)}

\title{Modelling the 1.8 \MeV\ Sky: Tests for Spiral Structure}

\author{J.~Kn\"odlseder$^{1,5,6}$, N.~Prantzos$^5$, K.~Bennett$^4$,
        H.~Bloemen$^2$, R.~Diehl$^1$, W.~Hermsen$^2$, U.~Oberlack$^1$,
	    J.~Ryan$^3$, and V.~Sch\"onfelder$^1$}

\institute{$^1$Max-Planck-Institut f\"ur extraterrestrische Physik,
	   Postfach 1603, 85740 Garching, Germany\\
	   $^2$SRON-Utrecht, Sorbonnelaan 2, 3584 CA Utrecht, The
	   Netherlands\\
	   $^3$Space Science Center, University of New Hampshire, Durham
	   NH 03824, U.S.A.\\
	   $^4$Astrophysics Division, ESTEC, ESA, 2200 AG
	   Noordwijk, The Netherlands\\
       $^5$Institut d'Astrophysique de Paris, 98bis Bd. Arago,
       75014 Paris, France\\
       $^6$Centre d'Etude Spatiale des Rayonnements, CNRS/UPS, BP 4346,
	   31029 Toulouse Cedex, France}

\offprints{J\"urgen Kn\"odlseder (Toulouse)}

\date{Received October 1995; accepted October 1995}

\maketitle
\markboth{J. Kn\"odlseder \etal:
	  Modelling the 1.8 \MeV\ Sky: Tests for Spiral Structure}
	 {J. Kn\"odlseder \etal:
	  Modelling the 1.8 \MeV\ Sky: Tests for Spiral Structure}

\begin{abstract}

COMPTEL imaging analysis revealed a patchy, asymmetric distribution of
diffuse 1.8 \MeV\ emission along the Galactic plane, which is
attributed to the decay of radioactive \al26\ in the ISM.
If massive stars were the major source of Galactic \al26, the 1.8 \MeV\
emission should be asymmetric and trace the spiral arms of the Galaxy,
presumed site of massive star formation.
Using model fits, we indeed find weak evidence in the COMPTEL data that the
observed 1.8 \MeV\ emission is at least partly confined to spiral arms.
We derive a total Galactic \al26\ mass of 2.5 \Msol\ from which at least
0.7 \Msol\ can be attributed to massive stars.

\end{abstract}

\section{Introduction}

Since the discovery of the 1.8 \MeV\ gamma-ray line emission from
radioactive \al26\ by Mahoney \etal\ (1982), the questions of its
origin and its distribution along the Galactic plane stimulated a
wave of research (see review of \cite{rf:pd95}).
Core collapse supernovae (SNe), Wolf-Rayet (WR) stars,
asymptotic giant-branch (AGB) stars, and O-Ne-Mg novae were suggested
as possible sites of significant \al26\ creation.
Early works assumed that the large mean lifetime
of $\tau_{26}\sim10^6$ yr and the low \al26\ yield per source will
lead to a smooth and symmetric distribution of 1.8 \MeV\ emission.
Prantzos (1991) was the first who dropped the assumption of
an axisymmetric source distribution in the Galactic plane
if massive stars were the dominant \al26\ producers.
He argued that star formation occurs predominantly inside the
spiral arms, especially in the case of massive stars.
Thus the 1.8 \MeV\ emission profile should reflect the structure
of these arms which is thought to be asymmetric with respect
to the Galactic centre.
Previous analysis of COMPTEL data indeed revealed an
asymmetry with more 1.8 \MeV\ emission from the southern
(Galactic longitude $l$=180\deg-360\deg) than from the northern
($l$=0\deg-180\deg) Galaxy (\cite{rf:diehl95}).
Additionally, the 1.8 \MeV\ sky map shows lumpy emission and
`hot spots'.
Prantzos (1993) noted that some emission maxima coincide with the
assumed tangential directions of Galactic spiral arms.
Thus a detailed study of the spiral arm hypothesis is of interest.
In this paper we will report on a comparison of COMPTEL phase I+II
data (May 1991 - August 1993) to models of Galactic \al26\
distribution with special emphasis on spiral structure.

\section{Instrument and Data Analysis}

COMPTEL has an energy resolution of $\sim8\%$ (FWHM) at 1.8 \MeV\ and
an angular resolution of 3.8\deg\ (FWHM) within a wide field of view
of about 1 steradian.
\gray\ photons are measured by their consecutive interactions
in two parallel detector planes where an incident photon is first
Compton scattered in the upper layer and then absorbed (although
often not completely) in the lower layer.
A detailed description of the instrument can be
found in Sch\"onfelder \etal\ (1993).
We analyzed the COMPTEL data in the three-dimensional imaging data-space
which is spanned by the scatter direction ($\chi$,$\psi$) and the
Compton scatter angle \phibar\ of the incident photons.
For the \al26\ study, we applied a 200 \keV\ wide energy window, centred
on 1.8 \MeV, to the data.
This encloses $48\%$ of all detected events from a celestial 1.809
\MeV\ source.

The Galactic models are represented as \al26\ source density functions.
1.8 \MeV\ model intensity maps are evaluated by integration of these
functions for a grid of Galactic longitudes $l$ and latitudes $b$ along the
corresponding lines of sight (\cite{rf:pd95}).
The convolution of these maps with the instrumental point-spread-function
leads to model distributions of 1.8 \MeV\ source events in the data-space.
By maximization of the overall data-space likelihood, the source models
are fitted along with a model for the instrumental background to the
data.
The background model was derived using independent measurements at
adjacent energy bands.
Data-space analysis in this approach suppresses continuum emission and
reveals only the sources of pure 1.8 \MeV\ line emission (\cite{rf:knoedl95}).
To reduce systematic uncertainties, the unknown \phibar-profile of the
background model was adjusted by the fit.

We tried to eliminate impacts of possible local 1.8 \MeV\ foreground
emission from our Galaxy-wide study:
for the observed 1.8 \MeV\ emission in Vela, probably associated with
the Vela SNR (\cite{rf:oberlack94}, \cite{rf:diehl95b}), and in Cygnus,
probably related to some nearby star forming regions (\cite{rf:delrio94}),
two source components with free intensity were included in the
background model.
Both were chosen to have uniform intensity within a circle of 10\deg\
in radius, centred on \lb=(259\deg, 0\deg) for Vela and on
\lb=(83\deg, 0\deg) for Cygnus.
Also, the region of the outer Galaxy between $l$=120\deg\ and $l$=240\deg\
was excluded from the analysis, because the 1.8 \MeV\ sky-map shows
significant emission near the anticentre which is probably due to nearby
\al26\ sources implied by its wide latitude extension.

The maximum likelihood technique (de Boer \etal\ 1992) was applied
to determine the parameters and the significance of the source models.
It uses the parameter $-2\ln\lambda$ to quantify the model-data
agreement, where $\lambda$ is the maximum likelihood ratio
$L$(background)/$L$(source + background).
Higher $-2\ln\lambda$ values signal a better fit of the model to the data.
Formally, $-2\ln\lambda$ obeys a $\chi^2_n$ probability
distribution, where $n$ is the number of free parameters of the
source model.
Roughly, the significance of the source model over
background follows $\sqrt{-2\ln\lambda}$, which is exact for $n=1$.
We visualized the model-data agreement for each fit by a longitude scan
of the data-space for measurement and fitted model using the software
collimation technique (\cite{rf:diehl93}).
An acceptance circle of 3\deg\ was selected which implies an effective
angular resolution of 10\deg-12\deg\ for the scans.

\section{Axisymmetric models}

\subsection{Tracers of Galactic \al26}

We start our discussion with axisymmetric models for which the source
density function only depends on the galactocentric radius $R$ and the
distance $z$ from the Galactic plane.
For the vertical profile we assumed throughout a uniform
exponential law with scale height $z_0$.
Unfortunately, the Galactic distribution of all \al26\ candidate sources
is poorly known because of the combined effects of visual obscuration,
uncertain sample completeness, and small-number statistics.
Consequently, we model their spatial distribution by tracers observed
either in our Galaxy or in external spiral galaxies.
Many such tracers have been proposed up to now, and we discuss only a
few that seem to be the most appropriate (see also Diehl et al., this volume).
For this purpose we divide the candidate sources in two classes:

\noindent {\bf Young population:}
Stars younger than $\sim10^8$ yr are classified as extreme Population I
objects.
{}From the \al26\ candidate sources, massive AGB stars, SNe, and WR
stars fall in this group.
The Galactic distribution of these objects has a small scale height
($z_0\sim90$ pc) and can be traced either by giant \HII\ regions or
giant molecular clouds (GMCs).
The distribution of GMCs is generally inferred from radio observations
of the CO $J=1\to0$ rotational transition at a wavelength of 2.6 mm.
Under the simplifying assumption that the \al26\ emissivity
is proportional to the molecular gas mass surface density we used
the radial $H_2$ distribution given in Fig. 1 of Dame (1993).
We added a Galactic centre flat disk component with radius of 500~pc
to account for the nuclear disk which is present in all CO surveys.
The total mass of this component was left as free parameter,
because the star-formation efficiency and therefore the \al26\ yield
in the Galactic centre might differ from that in the Galactic disk.

\noindent {\bf Intermediate population:}
Low-mass AGB stars (M$<2$\Msol) and novae have ages greater than a few
$10^9$ yr.
It is believed that these objects follow the luminosity profile of the
Galaxy which is composed of a disk and a bulge component (\cite{rf:bs80}).
The disk is assumed to be of the exponential form
$\sigma(R)\propto\exp(-R/R_0)$, where $\sigma(R)$ is the galactocentric
surface density and $R_0$ is the radial scale length of the disk.
Estimates for $R_0$ have been derived by numerous workers and span the
enormous range of 1 to 6 kpc (\cite{rf:wainscoat92}, \cite{rf:kdf91}).
Patterson (1984) found a disk scale height of $\sim180$ pc for the
intermediate population.
The bulge is commonly described as an oblate spheroid with an exponential
density decrease.
Wainscoat \etal\ (1992) fitted {\em IRAS} data using
$\rho(R,z)\propto x^{-1.8}{\rm e}^{-x^3}$, where
$x=\sqrt{R^2 + k_{\rm e}^2 z^2}/R_{\rm e}$, $k_{\rm e}=1.6$ and
$R_{\rm e}=2.0$ kpc are the axis-ratio and effective radius, respectively.
However, the relevance of the bulge component for \al26\ sources is
questionable because it possibly contains only objects older than $10^{10}$
yr, hence no \al26\ producing AGB stars or O-Ne-Mg novae.

\subsection{Results and discussion}

{}From the fit of an exponential disk we found an optimum scale length of
$5^{+5}_{-2}$ kpc and scale height of $180^{+240}_{-130}$ pc for the
Galactic 1.8 \MeV\ emission (all quoted uncertainties are statistical
$2\sigma$ errors).
The scale length is on the upper side of estimates at other wavelengths.
The scale height is only weakly constrained because of the poor angular
resolution of COMPTEL, but it overlaps with estimates which reach from
90 pc to 325 pc for the youngest and oldest stellar populations,
respectively (\cite{rf:bs80}).
For comparison of the data with the intermediate stellar population, we fixed
the scale height to 180 pc and added the bulge as additional component.
The inclusion of the bulge gave only marginal, insignificant,
fit improvement: there is no evidence for a bulge on top of an
exponential disk in our 1.8 \MeV\ COMPTEL data (Table \ref{tb:symmetric}).
To test for the young stellar population we fitted CO models with and
without a nuclear disk in the Galactic centre (GC) to the data.
For both cases the optimum scale height was $110^{+100}_{-70}$ pc which
is consistent with the scale height of the molecular gas.
We find preference for the model with the nuclear disk component at the
3$\sigma$ significance level, but the likelihood ratios for both CO models are
slightly worse than that of the `best disk'.

\begin{figure}
 \setlength{\unitlength}{1cm}
 \begin{minipage}[t]{8.8cm}           
  \begin{picture}(8.8,5.5)            
   \put(1.0,0.0){\makebox(8.8,5.5){
    \psfig{file=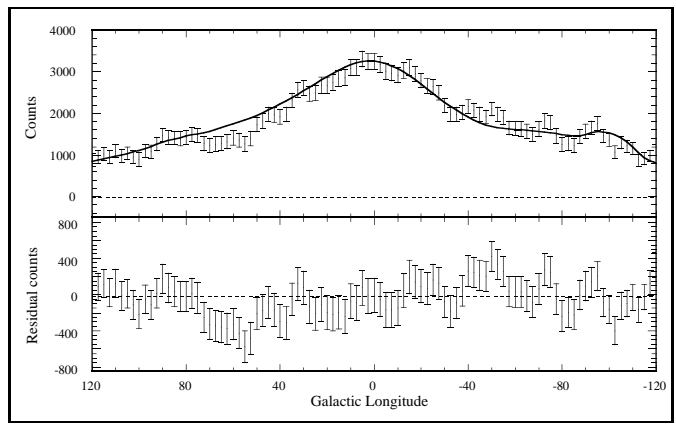,width=8.8cm,height=5.5cm}
    }}
   \put(0,0){\framebox(8.8,5.5)}
  \end{picture}
  \caption{\label{fig:axiquality}
	   Longitude scan derived by software collimation for the `best disk'
	   fit.
	   The upper panel shows the background subtracted profile, in the
	   lower panel the residual counts are plotted (observed-predicted).
	   Note, that we directly compare measured to predicted counts,
	   thus an axisymmetric distribution must not lead to a symmetric
	   longitude profile because of exposure variations along the
	   Galactic plane.}
 \end{minipage}
\end{figure}

As an illustration for axisymmetric models the longitude scan of the
`best disk' fit, derived by software collimation, is shown in
Fig.~\ref{fig:axiquality}
(note, that this scan technique has only poor angular resolution
(10\deg-12\deg), hence sharp features in the data are smeared out,
large-scale trends are emphasized at the cost of not showing small-scale fit
inadequacies).
Obviously, the `best disk' already gives a reasonable first-order
description of the 1.8 \MeV\ data.
However, there are some significant discrepancies between the model and
the data.
While there is a lack of counts in the northern Galaxy between
$l\approx50\dg - 75\dg$, excess counts are found around $l\approx-50\dg$
and $l\approx-75\dg$ in the southern Galaxy.
It is obvious that this north-south asymmetry cannot be explained
by any axisymmetric model.
Actually, the `best disk' is the optimal balance of this asymmetry which
explains why it yields the best likelihood ratio of all axisymmetric
models.
Therefore, we conclude that objects with a smooth, symmetric distribution
(as expected for low-mass AGB stars or novae) cannot be the solely source
of \al26\ in the Galaxy.

\begin{table}
\caption[]{Fit results for axisymmetric models. The last two columns
	   contain the total Galactic \al26\ mass in the disk and in a
	   possible Galactic centre (GC) component.
	   We quote $2\sigma$ statistical errors for the mass which
	   generally increase with the number of free model parameters.}
\begin{flushleft}
\begin{tabular}{lccrcc}
\noalign{\smallskip}
\hline
\noalign{\smallskip}
      & $R_0$ & $z_0$ &                & Disk  & Centre \\
Model & \multicolumn{1}{c}{(kpc)} & (pc)  & $-2\ln\lambda$ &
		 \multicolumn{2}{c}{\al26\ mass (\Msol)} \\
\hline
Best disk & $5^\ast$ & $180^\ast$ & 418.0 & $3.2\pm1.1$ &     -         \\
+bulge    & $6^\ast$ & 180        & 418.1 & $3.3\pm0.5$ & $0.07\pm0.19$ \\
CO        &    -     & $110^\ast$ & 403.8 & $2.4\pm0.5$ &     -         \\
CO+GC     &    -     & $110^\ast$ & 412.3 & $2.2\pm0.5$ & $0.17\pm0.12$ \\
\noalign{\smallskip}
\hline
\noalign{\smallskip}
\end{tabular}
\\
{\footnotesize
 $^\ast$ parameters optimized by the fit}
\label{tb:symmetric}
\end{flushleft}
\end{table}

\section{Spiral models}

We now drop the assumption of an axisymmetric source distribution
and investigate the hypothesis that the Galactic 1.8 \MeV\
emission is correlated with spiral arms.
Unfortunately, the spiral structure of the Galaxy is not well
established and even the number of spiral arms is under debate
(e.g. \cite{rf:elmegreen85}).
The most reliable large-scale picture of the Galactic spiral
structure is probably deduced from the distribution of giant \HII\
regions.
Their distances can be estimated spectro-photometrically from the distances
to the exciting O/B-stars and do not depend on models of Galactic rotation
(in contrast to spiral structure derived from \HI\ or CO surveys).
We adopt the spiral model of Taylor \& Cordes (1993) who adjusted a
four-arm spiral pattern based on giant \HII\ regions and radio-survey
tangent points to pulsar dispersion measures and
interstellar scattering measurements.
Their aim was to obtain a quantitative model for the distribution of
free electrons in the Galaxy to estimate pulsar distances from dispersion
measures.
Recently, Chen \etal\ (1995) pointed out that free electrons
could be a valuable tracer of \al26\ because they are mainly produced
by the massive star population.
Therefore we directly compare the Taylor \& Cordes (TC) free-electron model
to the COMPTEL 1.8 \MeV\ data.
We tentatively added a nuclear disk component (c.f. section 3.1) of 90 pc
scale height in the Galactic centre region
where the TC model is only weakly constrained by the pulsar data.

\begin{figure}
 \setlength{\unitlength}{1cm}
 \begin{minipage}[t]{8.8cm}           
  \begin{picture}(8.8,5.5)            
   \put(1.0,0.0){\makebox(8.8,5.5)
    {
     \psfig{file=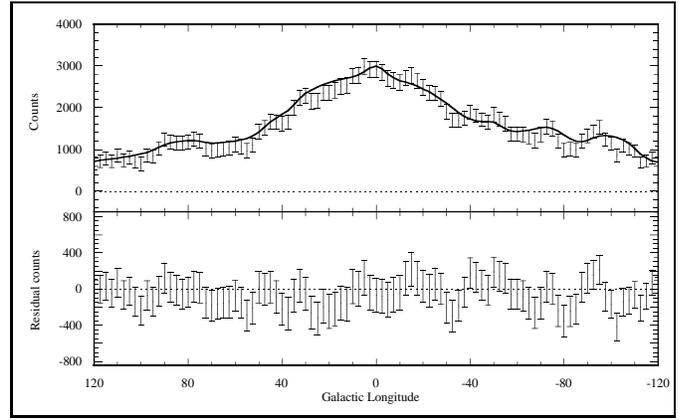,width=8.8cm,height=5.5cm}
     }}
   \put(0,0){\framebox(8.8,5.5)}
  \end{picture}
  \caption{\label{fig:spiral}
	   Longitude scan derived by software collimation for the
	   Taylor \& Cordes spiral model including a nuclear disk component.}
 \end{minipage}
\end{figure}

The likelihood ratio for both the TC model with and without the nuclear disk
component is better than that of the best exponential disk (see Table
\ref{tb:spiral}).
{}From the longitude scan (Fig. \ref{fig:spiral}) we see that the TC model
fits the north-south asymmetry better than the axisymmetric models.
Besides the global asymmetry, the TC model also explains the two excesses
at $l\approx-50\dg$ and $l\approx-75\dg$ due to the presence of spiral arm
tangent points in these directions.
Additional, but less outstanding tangent points of the model at
$l\approx-30\dg$, $l\approx30\dg$, and $l\approx50\dg$ are almost invisible
in the scan because they are smeared out due to the poor angular resolution
of the software collimation technique (see above).
However, except the $l\approx-30\dg$ tangent, all spiral arm tangent points
of the model coincide well with count excesses in the data supporting the
hypothesis that Galactic \al26\ is at least partly confined to spiral arms.

We also studied a more analytical model which consists of the TC spiral
pattern on top of an exponential disk. This model allows that some massive
stars lie outside the spiral arms, but would also be valid
if Galactic \al26\ has a composite low- and high-mass star origin.
{}From the fit we obtained an optimum disk scale length of
$3.5^{+2.5}_{-1.5}$ kpc and scale height of $180^{+240}_{-130}$ pc.
The likelihood ratio is similar to that of the TC+nuclear disk model,
thus the data cannot tell us which of the two models is a more reliable
representation of the Galactic \al26\ distribution.
The total \al26\ mass of 2.7 \Msol\ is comprised of $2.0\pm0.7$ \Msol\
for the disk and $0.7\pm0.3$ \Msol\ for the spiral arms.
Therefore, the total Galactic \al26\ mass created by massive stars is
bracketed by $\sim0.7$ \Msol\ from the arm component of the disk+arm model
and by $\sim2.5$ \Msol\ from the TC free electron model.

\begin{table}
\caption[]{Fit results for spiral models.}
\begin{tabular}{lccc}
\noalign{\smallskip}
\hline
\noalign{\smallskip}
      &                 & Total Galactic       & Nuclear disk \\
Model &  $-2\ln\lambda$ & \multicolumn{2}{c}{\al26\ mass (\Msol)}  \\
\hline
TC                   & 419.6 & $2.6\pm0.3$ & - \\
TC+nuclear disk      & 423.2 & $2.5\pm0.4$ & $0.11\pm0.06$ \\
Best disk+arms       & 423.7 & $2.7\pm0.8$ & - \\
\noalign{\smallskip}
\hline
\noalign{\smallskip}
\end{tabular}
\\
\label{tb:spiral}
\end{table}

\section{Conclusions}

We have compared the COMPTEL 1.8 \MeV\ data from observation phases I+II
to axisymmetric and spiral-arm models of Galactic \al26\ distribution.
All models were detected at a significance level of $>20\sigma$
above background.
To first order, the observed 1.8 \MeV\ emission is well represented by
axisymmetric models.
However, details of the data like the north-south asymmetry and some regions
with significant count excesses are better described by models which
incorporate the Galactic spiral structure.
Our best fit model holds a total Galactic \al26\ mass of $\sim2.5$ \Msol\
from which at least $0.7$ \Msol\ are produced by massive stars.
Thus, massive stars clearly contribute to the observed \al26\ in
the Galaxy but we can certainly not exclude from this work that a large
fraction of \al26\ is produced by low-mass AGB stars or novae.

\begin{acknowledgements}
The COMPTEL project is supported by the German government through
DARA grant 50 QV 90968, by NASA under contract NAS5-26645, and by
the Netherlands Organisation for Scientific Research NWO.
\end{acknowledgements}


\end{document}